\documentclass[aps,twocolumn,amssymb]{revtex4}
\usepackage{epsfig,amsmath}
\usepackage{graphicx}
\usepackage{hyperref}
\usepackage{url}

\def\beq{\begin{equation}}
\def\eeq{\end{equation}}
\def\imagetop#1{\vtop{\null\hbox{#1}}}

\begin{document}

\title{Shape-shifting droplet networks}

\author{T. Zhang, Duanduan Wan, J. M. Schwarz, and M. J. Bowick}
\affiliation{Department of Physics, Syracuse University, Syracuse, NY 13244, USA}

\begin{abstract}
{We consider a three-dimensional network of aqueous droplets joined by single lipid bilayers to form a cohesive, tissue-like material. The droplets in these networks can be programmed to have distinct osmolarities so that osmotic gradients generate internal stresses via local fluid flows to cause the network to change shape. We discover, using molecular dynamics simulations, a reversible folding-unfolding process by adding an osmotic interaction with the surrounding environment which necessarily evolves dynamically as the shape of the network changes.  This discovery is the next important step towards osmotic robotics in this system. We also explore analytically and numerically how the networks become faceted via buckling and how quasi-one-dimensional networks become three-dimensional.} 
\end{abstract}
\maketitle

In nature there are many biomaterials that are internally programmed to morph into complex structures that actively and adaptively interact with the environment. Polypeptide chains fold into proteins and tubular lipid membranes self-assemble in branching networks to form the endoplasmic reticulum. Organisms as a whole develop shape by reorganizing the spatial distribution of their constituent cells in morphogenesis. 

These biological examples have inspired the development of {\em programmable} materials that controllably fold into designated structures. At the nanometer scale, the programmable chemistry of Watson-Crick base pairing allows DNA to self-assemble into a tetrahedron~\cite{sadowski2014developmental}. At the millimeter scale, efficient algorithms have been constructed to generate self-folding three-dimensional polyhedra from two-dimensional nets, driven by the minimization of the surface tension of liquid hinges that either rotate or fuse panels into place~\cite{pandey2011algorithmic}. At the centimeter scale, researchers have created a self-folding robot that goes from flat to walking in several minutes without external intervention~\cite{felton2014method}. 

\begin{figure}[h]
\begin{center}
\includegraphics[width=8.0cm]{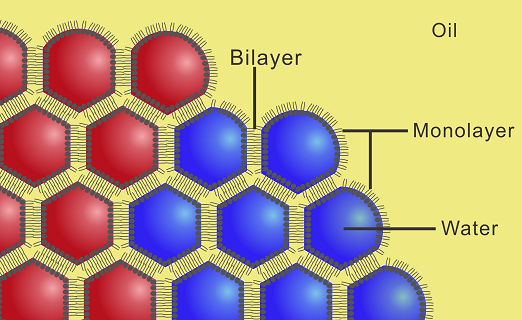}
\caption{(Color online) Schematic image showing droplets of different aqueous solutions printed into a solution of lipids in oil. The droplets acquire a lipid monolayer and form bilayers with droplets in the developing network.  }
\label{Bayley}
\end{center}
\end{figure}

A beautiful realization of these ideas comes from the Bayley group who print tens of thousands of micron-sized aqueous droplets each joined by single lipid bilayers~\cite{Bayley, doi:10.1021/la9801413, doi:10.1021/ac0613479, doi:10.1021/nl0611034, doi:10.1021/ja072292a} to form a cohesive, tissue-like material (Fig.~\ref{Bayley}). The droplets in these networks can be endowed with different osmolarities. The resultant osmotic pressure leads to local fluid flow from low to high concentration. This swells the high concentrations droplets and shrinks the low concentration droplets, leading to internal stresses which distort the shape of the  network in specific ways depending on the initial geometry of the network and the concentration differences. The initial droplet network experiments explored the formation of a hollow sphere from an initial two-dimensional four-petal-shaped structure (see \href{http://syrsoftmatter.syr.edu/wp-content/uploads/2015/05/Movie-1-folding_flower.mp4}{SM Movie-1}), thus demonstrating the spontaneous assembly of three-dimensional shapes.

{To push the capability of this system into the realm of osmotic robotics, the droplet network must be able to make {\it reversible} shape changes. We investigate this possibility by turning to programmable interactions with the network's environment.  Past examples of a system's environment driving shape change do exist.} A strip of paper will, for instance, spontaneously curl up in your hand due to the interaction between the paper and the moist evaporative boundary layer of the hand~\cite{maha1}. Bacillus spores respond to relative changes in environmental humidity with low humidity causing the spores to shrink and high humidity causing the spores to expand~\cite{maha2}. When these spores self-assemble into a dense monolayer sitting on a substrate, one can cycle the relative humidity of the environment to form an actuator. Osmotic actuation is used in plants, perhaps because it can generate a variety of plant movements, depending on the environmental conditions~\cite{plants}, without consuming much power. 

{Here we use predominantly computational methods to explore this plant analogue in the synthetic, micron-sized droplet network and show how shape shifting can be made reversible by exploiting the dynamically evolving coupling to the environment, which here is simply the surrounding solvent. We illustrate reversibility with shape shifting from a four-petal configuration to a hollow sphere transition and back. We also demonstrate the generalizability of this property by introducing first the formation of tetrahedral shells via folding and then a reversible tetrahedral shell folding-unfolding-folding process. Tetrahedra, the simplest and least symmetric class of the regular polyhedra, can serve as mesoscopic building blocks for molecules and bulk materials with $sp^3$-like directional bonding~\cite{nelson2002toward, glotzer2007anisotropy}. Our reversibility findings open the door to osmotically-driven small scale robotics.

Finally, to better determine the range of shape formation in these droplet networks, we revisit the formation of rings studied in Ref.~\cite{Bayley} and identify a new buckling transition from a round to a polyhedral ring after ring closure. This configuration also has the advantage of being analytically tractable. We also explore spiral formation as they provide an interesting example of an initial quasi-one-dimensional structure generating a three-dimensional structure to expand the self-assembly capability of the system.}

{\it Model--} We model the droplet network using molecular dynamics in three-dimensional Euclidean space. {Each droplet $i$ is treated as a point mass $m_{i}$, with an associated radius $R_{i}$} and osmolarity (defined as the number of osmoles (Osm) of solute per litre (L) of solution) $C_i$. As described in Ref.~\cite{Bayley}, each droplet interacts with neighboring droplets via an elastic interaction and an osmotic interaction. The elastic interaction potential of a pair of droplets $i$ and $j$ via the bilayer attaching the two droplets is written as
\begin{equation}
E_{ij}=
\left\{
\begin{aligned}
&\frac{1}{2}k \left[ r_{ij}-l\left(R_i + R_j\right) \right]^2~~~~~~~{\rm for}~r_{ij} \leq R_i + R_j\\
&0~~~~~~~~~~~~~~~~~~~~~~~~~~~~~~~~~~{\rm for} ~r_{ij} > R_i + R_j\\
\end{aligned}
\right.
\end{equation}
where $k$ is the spring constant, $r_{ij}$ is the distance between a pair of droplets $i$ and $j$, 
and $l=0.8$ represents the change of the equilibrium length due to the deformation of two droplets when they are fused~\cite{Bayley}. A  damping force for each droplet proportional to its velocity is included as demanded by Stokesian flow. The exchange of water between droplets of different osmolarities is described by Fick's first law as
\begin{equation}
\label{equation_water_exchange}
J_{ij}=A_{ij} D \left( C_j - C_i \right), 
\end{equation}
where $J_{ij}$ is the volume of water transferred per unit time from a droplet $i$ with osmolarity $C_i$ to a connected droplet $j$ with osmolarity $C_j$, $A_{ij}$ is the common interfacial area of two connected droplets $i$ and $j$ and $D$ is an effective permeability coefficient taken to be constant. The flow of water between two connected droplets changes the size of each droplet.
To simulate the dynamics of the droplet network, the net force on each droplet $\vec{F}_i$ is described by
\begin{equation}
\vec{F}_i=-\sum_{<ij>}\frac{dE_{ij}}{dr_{ij}}\hat{r}_{ij}-\gamma\vec{v}_i=m\frac{d^2\vec{r}_i}{dt^2},
\end{equation}
where $\vec{v}_i$ is the velocity of droplet $i$ and $\gamma$ is the damping coefficient. {In experiments, the radius of the droplet is of order tens of microns, the spring constant is of order a few milliNewtons per meter and the time scales for mechanical relaxation and water transfer are seconds and tens of minutes, respectively.} We invoke this approach because the time scale for mechanical equilibrium is much faster than the time scale for water transfer. More simulation details may be found in SM II. 

\begin{figure}[h]
\centering
\begin{tabular}[t]{|c|c|c|}
\hline
\imagetop{\includegraphics[width=2.5cm]{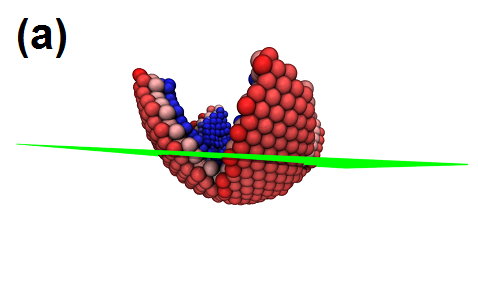}} &
\imagetop{\includegraphics[width=2.5cm]{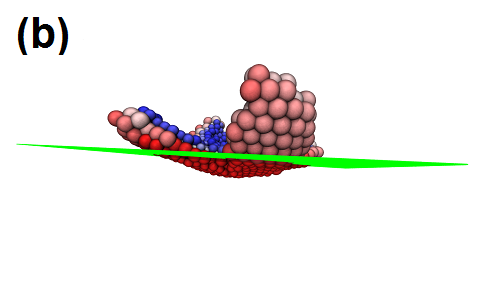}} &
\imagetop{\includegraphics[width=2.5cm]{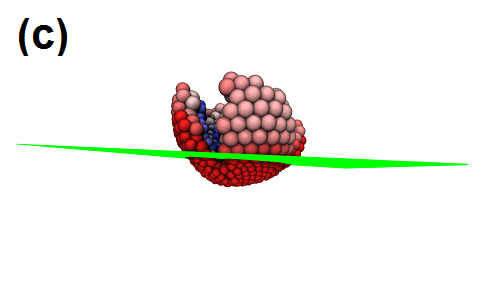}} \\
\hline
\end{tabular}
\caption{(Color online) (a)-(c) Three snapshots from the reversible folding process. See \href{http://syrsoftmatter.syr.edu/wp-content/uploads/2015/05/Movie-5-folding_unfolding_refolding.mp4}{SM Movie-5} for the full sequence of snapshots.}
\label{unfold}
\end{figure}

\medskip
{\it Reversible Folding--} {Let us start with the four-petal-
shaped structure studied experimentally in Ref.~\cite{Bayley}.} By adding an osmotic interaction with the environment we can realize a {\em reversible} folding-unfolding process, as shown in Fig.~\ref{unfold} (see \href{http://syrsoftmatter.syr.edu/wp-content/uploads/2015/05/Movie-5-folding_unfolding_refolding.mp4}{SM Movie-5}). 
To generate reversible folding we place part of the folded droplet network into a medium with higher osmolarity $C_m$ ($C_m>C_{1,2}$) so that water will flow from the droplets on the bottom layer of the ``flower" to the medium. More precisely, the bottom layer droplets located below a horizontal $x-y$ plane at $z=h$ are exposed to the higher osmolarity medium. As the flower folds, an increasing number of bottom layer droplets naturally become exposed to the medium. They therefore lose water, to the surrounding medium, and start to unfold again. The overall volume of the unfolded flower also drops, preventing a complete reversal. As the bottom droplets continue losing water to the medium, the top droplets are also losing water to the bottom droplets.

The detailed evolution depends on the the osmolarity difference and the total contact area through which the water is passing. Lower $h$, which means the smaller the contact area between the bottom droplets and the medium, and the smaller $C_m$, which means a smaller osmolarity difference, will lead to a slower rate of water transfer between the bottom droplets and the medium. For top layer osmolarity $C_{top}=0.1$ and bottom layer osmolarity $C_{bottom}=1.0$, one finds that the top layer loses more water to the bottom layer than the bottom layer itself loses to the surrounding medium.  This eventually reverses the unfolding and the structure starts to fold once again. As discussed in SM II, our simulation uses the same algorithm as before, other than a modification of Fick's first law (Eq. (2)) resulting from  the separation of time scales between mechanical relaxation and water transfer. 

To characterize the reversibility of the system, we define reversibility as 
\begin{equation}
\rho=\left( D_{max\_before} + D_{max\_after} - 2D_{min}\right) / K_{norm},
\end{equation}
where $D_{min}$ is the minimum depth of unfolding,  $D_{max\_before}$ is the maximum depth of the flower before the ``flower" unfolds, $D_{max\_after}$ is the maximum depth of the flower after it reaches a minimum and refolds, 
and $K_{norm}$ is a normalization constant. For $\rho=1$, the system is fully reversible in the sense that the flower opens out fully before folding back up again, while for $\rho=0$ there is no reversal.

\begin{figure}[h]
\begin{center}
\includegraphics[width=8.0cm]{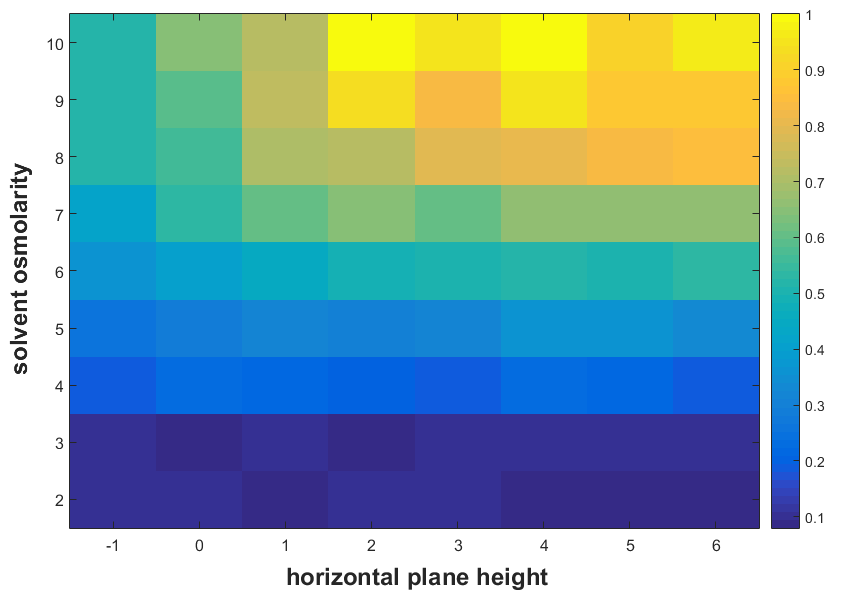}
\caption{(Color online) Reversibility dependence on the $z$ coordinate of the horizontal plane and the osmolarity of the surrounding medium.}
\label{reversibility}
\end{center}
\end{figure}

Fig.~\ref{reversibility} shows the dependence of $\rho$ on the horizontal plane's $z=h$ coordinate and the media's osmolarity $C_m$. As the horizontal plane rises, the reversibility increases since there are more bottom droplets in contact with the medium. The reversibility also increases with a rise in the media's osmolarity $C_m$. 

\begin{figure}[h]
\centering
\begin{tabular}[t]{|c|c|c|}
\hline
\imagetop{\includegraphics[width=4.0cm]{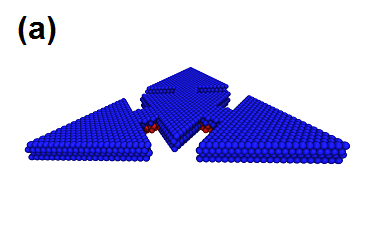}} &
\imagetop{\includegraphics[width=4.0cm]{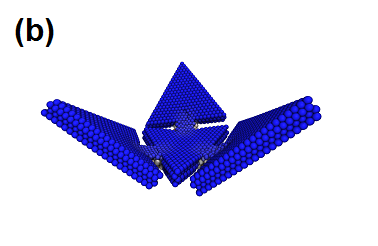}} \\ \hline
\imagetop{\includegraphics[width=4.0cm]{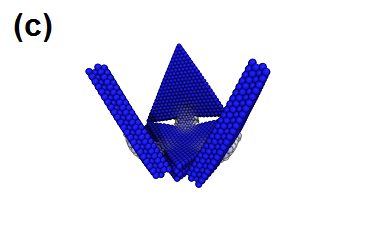}} &
\imagetop{\includegraphics[width=4.0cm]{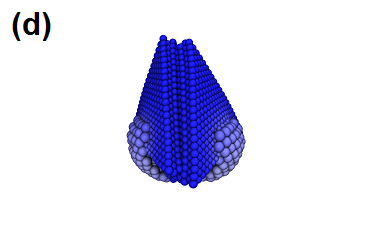}} \\ 
\hline
\end{tabular}
\caption{(Color online) (a)-(d) Forming a tetrahedral shell ($C_1=1.0$ and $C_2=5.0$). The color represents osmolarity with blue/red representing low/high osmolarity with white intermediate. }
\label{tetrahedron}
\end{figure}

{We can generalize this notion of reversibility to tetrahedral shells folding and unfolding.  Before addressing the reversibility, we first demonstrate how a tetrahedral shell can form.} We choose as initial droplet network a central triangle connected to three other triangles by hinges on each of its sides. This is an example of a two-dimensional net of polygons and hinges (creases in origami) that can be folded into three-dimensional structures. We use four layers of droplets, all of osmolarity $C_1$, to create  sufficiently rigid faces. For each hinge, the top two layers of droplets have osmolarity $C_1$, while the bottom two layers droplets have osmolarity $C_2>C_1$. The flow of water to the outer layers causes the hinges to bend upward, closing all faces into a tetrahedral shell (see Fig.~\ref{tetrahedron} and SM Movie-4). 
The osmolarity difference must be tuned to achieve tetrahedral folding. If $C_1=1.0$ and $C_2<4.0$, the osmolarity difference is too small to drive tetrahedral closure. {This threshold can be approximated analytically. See SM III for details.} The critical osmolarity difference may be lowered or raised by using wider or thinner hinges respectively.

{Since this structure has yet to be explored experimentally, we also test for robustness of the tetrahedral shell formation to disorder. We do so by randomly diluting droplets in the initial state. For 20 percent dilution and lower, tetrahedral shell formation persists.  This percentage can be increased as the width of the hinges increases. Moreover, should we allow for one dilution percentage in the hinges and a second, independent dilution percentage in the triangles, we expect the percentage of dilution in the triangles at which the tetrahedral shell formation is no longer robust to be higher.}

{And now for the reversibility capability of the tetrahedral shell. See \href{http://syrsoftmatter.syr.edu/wp-content/uploads/2015/12/movie-6-folding_unfolding_refolding_tetrahedron.mp4}{SM Movie-6}. If we assume that only the bottom layer droplets of the hinge transfers water with the solvent, there is a smaller window of the osmolarity of the solvent $C_m$ than the four-petal-shaped structure for the reversible folding to occur. In particular, the relation between reversibility capability and $C_m$ is nonmonotonic, e.g. for small and large values of the osmolarity of solvent, the folding is not reversible.  For the larger values of $C_m$, the triangles in the tetrahedral shell will not be flush as the volume of the hinge decreases faster than the triangle. Moreover, the reversibility depends on the height of the surface of the solvent strongly than the four-petal-shaped structure because the hinge is shorter than each petal of the ``flower''.\\}

\begin{figure}
\centering
\includegraphics[width=3.5in]{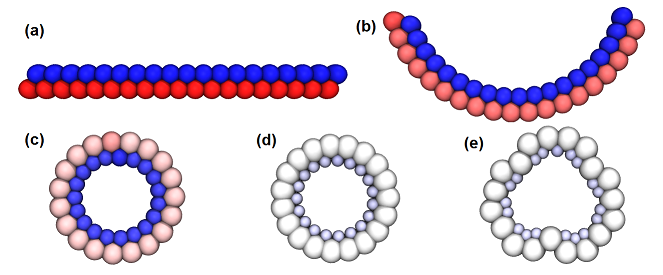}
\caption{(Color online) (a)-(e) Buckling of a ring with $N=38$ total droplets. The initial osmolarities of the two rows are $C_1=0.1$ (blue) and $C_2=1.0$ (red), respectively. Snapshots were generated using the Visual Molecular Dynamics (VMD) package \cite{HUMP96} and rendered using the Tachyon ray tracer \cite{STON1998}. }
\label{buckling}
\end{figure}

{\it Rings and buckling--} The buckling of a circular elastic ring subject to an external radial pressure has been extensively studied in applied mechanics \cite{BIEZENO1948}. {This buckling can presumably occur due to internal pressures as well. To study this possibility} in the context of droplet networks we evolve from an initial configuration consisting of two rows with different osmolarities (see \href{http://syrsoftmatter.syr.edu/wp-content/uploads/2015/05/Movie-2-folding_ring.mp4}{SM Movie-2}). FIG. \ref{buckling} shows ring closure for $N=38$ total droplets. The initial osmolarity of the top row is $C_1=0.1$ (blue) and that of the bottom row is $C_2=1.0$ (red). After each row closes to form a ring there is still an osmolarity difference between the outer and inner rings, as can be seen from Fig.\ref{buckling}(c). This residual osmolarity mismatch is followed by a ring buckling transition, as shown in Fig.\ref{buckling}(e).      

\begin{figure}[h]
\begin{center}
\includegraphics[width=8.0cm]{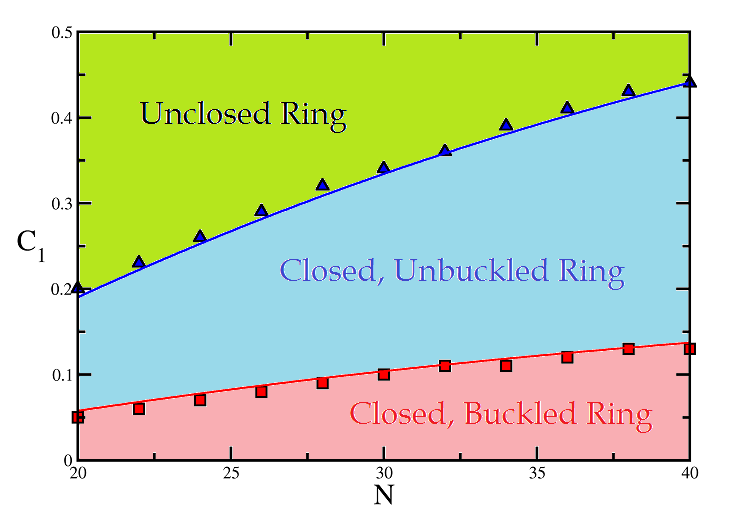}
\caption{(Color online) Plot of the ring buckling phase diagram as a function of top row osmolarity $C_1$ and total number of droplets $N$. The bottom row osmolarity is fixed at $C_2=1.0$. The symbols, obtained from simulation, should be compared with lines showing analytical results.}
\label{ring_phase}
\end{center}
\end{figure}

The final shape depends on the osmolarity difference and the number of droplets. In Fig.~\ref{ring_phase}, we sketch the phase diagram for ring closure and buckling as a function of the top row osmolarity $C_1$ and the total number of droplets $N$. The bottom row osmolarity is fixed at $C_2=1.0$. There are three phases. For a given value of $N$, rings do not form at all until the osmolarity difference $\Delta C = C_2 - C_1$ exceeds a threshold. The threshold value is computed analytically in SM {IV}. For a range of $\Delta C$ one then finds smooth rings. For yet larger $\Delta C$ the closed ring buckles. An upper bound for the buckling transition can also be obtained analytically (see SM {IV} ). The larger $N$, the easier it is to form a ring and so the smaller is the threshold osmolarity difference for ring closure and subsequent buckling. 
\begin{figure}[h]
\centering
\begin{tabular}[t]{|c|c|c|}
\hline
\imagetop{\includegraphics[width=6.0cm]{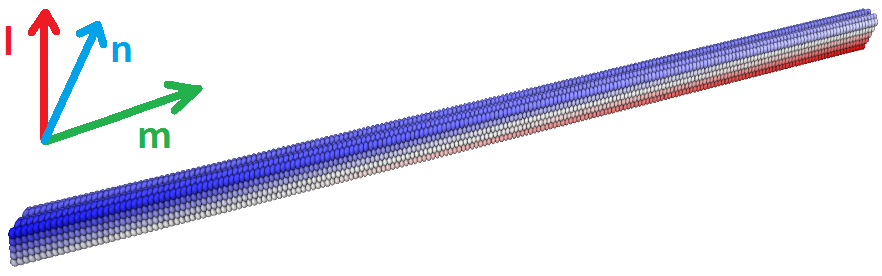}} &
\imagetop{\includegraphics[width=2.0cm]{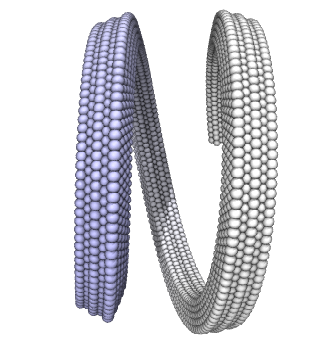}} \\ \hline
\end{tabular}
\caption{(Color online) Typical initial (left) and final (right) configurations for the formation of spirals.}
\label{spiral}
\end{figure}

{\it Spiral formation--} {Finally, to explore the possibility of generating three-dimensional structures from quasi-one-dimensional structures, we explore spiral formation.} We choose as initial state a $200\times5\times5$ rectangular slab (see Fig.~\ref{spiral} and \href{http://syrsoftmatter.syr.edu/wp-content/uploads/2015/12/Movie-3-folding_spiral.mp4}{SM Movie-3}). Each droplet can be indexed by integer orthogonal coordinates $l,m,n$, with $0\leq m < 200$, $0\leq n < 5$ and $0\leq l <5$. The initial osmolarity of each droplet is set to be 
\begin{equation}
C_{mnl}=\left( 1.0 - K_n\,n \right)\left( 1.0+K_l\,l\frac{(m-100)}{100} \right) \ ,
\end{equation}
where $K_n$ and $K_l$ are free parameters. We explore how spiral formation process depends on $K_l$ and $K_n$ for fixed slab size.

\begin{figure}[h]
\begin{center}
\includegraphics[width=8.0cm]{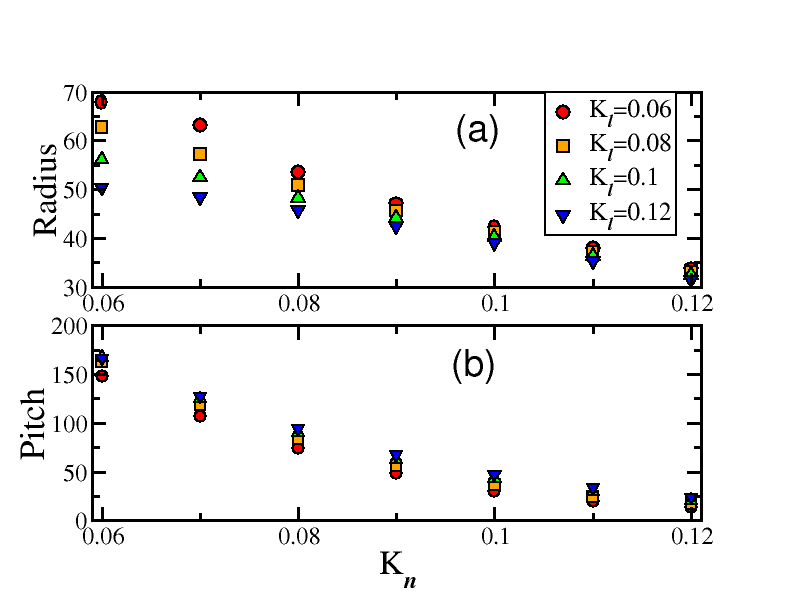}
\caption{(Color online) (a) The radius of the spiral as a function of $K_n$ for different $K_l$s.  (b) The pitch of the spiral as a function of $K_n$ for different $K_l$s. }
\label{radius}
\end{center}
\end{figure}

Fig.~\ref{radius}(a) shows how the radius of curvature of the spiral depends on $K_l$ and $K_n$. $K_n$ determines how efficiently the slab folds up in the $m-n$ plane. As $K_n$ increases the rod folds up more effectively and the radius of curvature decreases. A positive value of $K_l$ will make each part of the slab fold up differently in the $l$ direction. If $K_l=0$, for example, the two ends of the slab will fold and then meet each other, leading to a ring.
Note that the radius of the spiral also decreases with increasing $K_l$, making folding more efficient. 

Fig.~\ref{radius}(b) plots the pitch of the spiral as a function of $K_l$ and $K_n$. The pitch depends mainly on $K_n$,  
with more efficient folding occurring at large $K_n$. There is a slight dependence on $K_l$.

{\it Discussion---} {We have demonstrated for the first time how to, in principle, reverse the large-scale shape change in these droplet networks with the four-petaled flower folding, unfolding, and folding back again. We have also demonstrated the generalizability of this reversibility property to the folding-unfolding-refolding of tetrahedral shells.} {Forming a general shape from a net with folds in general (and the subsequent unfolding and re-folding) requires simultaneous fold movements. This should be possible to achieve with the right set of osmolarity parameters.} {The analytical analysis of the hinge in SM III also paves the way for analytically predictive estimates of the formation of more general shapes.} Our folding-unfolding-refolding processes capitalizes on the interaction between part of the droplet network and the surrounding medium (environment) and relies on the medium having an osmolarity larger than the osmolarities in the droplet network. Reversibility is a  first step towards osmotic robotics, namely a robotic gripper. The reversibility of our model is limited, however, because (1) the droplet network folds, unfolds, and folds again only once and (2) the volume of the final state of the folded shell (spherical or tetrahedral) is smaller than if it had just simply folded.   

To address the limited reversibility of our model, a recent study~\cite{ActiveOsmosis} extends osmosis to active solutes containing, for example, self-propelled colloids~\cite{golestanian2005propulsion, howse2007self, golestanian2007designing, sciortino2009phase} or hot nanoparticles~\cite{radunz2009hot, joly2011effective}. The study finds that active solute activity increases the osmotic pressure and can also expel solvent from the solution. By using active solutes, the solvent can be controlled to flow from the higher osmolarity to lower, which is reversed from the usual passive solute situation. The use of reverse osmosis via active solutes could make the folding-unfolding process completely reversible. Further study, however, is needed to assess the feasibility of active solutes to achieve such a goal.

We have also predicted the buckling of rings and the formation of spirals in these droplet networks driven by osmolarity gradients. For ring formation, we find a subsequent buckling transition for large enough osmolarity difference. For spiral formation, we determine how the size of the gradients in each of the three dimensions affects both the radius and pitch of the spiral. {Both the programmable buckling and the quasi-one-dimensional to three-dimensional structure formation enhance the programmable shape-shifting capability of the network. Moreover, robots could use buckling to generate sudden changes of geometry and spiral formation to enhance maneuverability.}

The authors acknowledge useful comments by M. C. Marchetti on an earlier draft of the manuscript. MB thanks L.~Mahadevan for stimulating discussions. This research was supported by the Soft Matter Program at Syracuse University.

\bibliography{References}

\newpage
\clearpage
\onecolumngrid
\begin{center}
\textbf{\large Supplementary Material for ``Shape-shifting droplet networks"}
\end{center}
\setcounter{equation}{0}
\setcounter{figure}{0}
\setcounter{table}{0}
\setcounter{page}{1}
\makeatletter
\renewcommand{\theequation}{S\arabic{equation}}
\renewcommand{\thefigure}{S\arabic{figure}}

\section{Selected Movies}
\begin{enumerate}
\renewcommand{\labelenumi}{\theenumi)}
\bfseries \item A four pedaled structure folds into a flower shaped hollow sphere: \\
\normalfont Movie-1-folding\_flower.mp4
\bfseries \item Ring formation and subsequent buckling: \\
\normalfont Movie-2-folding\_ring.mp4
\bfseries \item A 200x5x5 rectangular slab folds into a Spiral: \\
\normalfont Movie-3-folding\_spiral.mp4
\bfseries \item Tetrahetron formation: \\
\normalfont Movie-4-folding\_tetrahetron.mp4
\bfseries \item By adding an osmotic interaction with the environment, a folded flower shaped structure can unfold and then fold again: \\
\normalfont Movie-5-folding\_unfolding\_refolding.mp4
{\bfseries \item A tetrahedron can also fold, unfold, and fold again:\\
\normalfont Movie-6-folding\_unfolding\_refolding\_tetrahedron.mp4\\}
\end{enumerate}

\section{Further simulation details}

To simulate the dynamics of the droplet network, the net force on each droplet $\vec{F}_i$ is described by
\begin{equation}
\vec{F}_i=-\sum_{<ij>}\frac{dE_{ij}}{dr_{ij}}\hat{r}_{ij}-\gamma\vec{v}_i=m\frac{d^2\vec{r}_i}{dt^2},
\end{equation}
where $\vec{v}_i$ is the velocity of droplet $i$ and $\gamma$ is the damping coefficient. Following Ref.~\cite{Bayley}, we assume $m$, $k$, and $D$ are the same for all droplets. Because the time scale for mechanical relaxation is much faster than the time for water transfer (seconds compared to tens of minutes), any global shape change is in mechanical quasi-equilibrium. Thus the simplifying assumption of identical and constant $m$, $k$, and $D$ for all droplets should not affect our results in any significant way. We chose, in simulation units, $D=0.002$, $k=10^3$, and $m=0.2$. The value of the damping coefficient, $\gamma=1.1$, was chosen so that there are no oscillations between the droplets when they bind as observed experimentally.  The osmolarities ($C_i$) were chosen to give gradients similar to those for the flower-closing experiment. Length, time, and mass scales in simulation units can be converted to microns, seconds, and grams by matching to experiment. 

Implementing different initial osmolarity gradients and droplet configurations yields folding into a variety of important structures such as rings, spirals, and tetrahedra. To search for such structures, the droplets are initially positioned in hexagonal closed-packed arrangements {with a common initial volume $V_{0}=10$ ($V_{0}=\frac{4}{3}\pi R_{0}^{3}$, with $R_{0}$ the initial radius)} and equilibrated first without water exchange, after which osmosis is switched on. Any two droplets in contact are then connected via the elastic interaction. Once the elastic interaction is established, water is exchanged via Fick's first law and the radius of each droplet is updated accordingly { (we assume the droplets always keep a spherical shape)}. { The interfacial area $A_{ij}$ in Eq.~\ref{equation_water_exchange} in the main text is approximated as $A_{ij}=\pi R_{ij}^{2}$, where $R_{ij}= \mbox{min} \left\lbrace R_{i}, R_{j} \right\rbrace$.} The position of the center of each droplet is then updated using a fourth-order Runge-Kutta scheme to obtain the position of the centers at the subsequent time step with $\Delta t=0.01$ in simulation units. We have checked that our simulation results are robust to making the time step as small as $\Delta t=0.001$ and as large as $\Delta t=0.02$. As for the computational cost involved, a 4372 particle simulation for tetrahedron formation over 250,0000 simulation time steps took approximately 3 hours and 19 minutes on a computer with 2 quad core 2.66GHz processors. 

To model the droplet network osmotically interacting with the solvent in a controlled way, we place the folded droplet network in a medium with osmolarity $C_m$ exceeding any individual osmolarity of the droplets. The surrounding medium can only interact with part of the droplet network, as described above. We did not simulate the medium explicitly and neglected any mechanical response of the medium. { More simulation details and the experimental values of parameters can be found in Ref.~[4].}

{
\section{Details of an analytical analysis of each hinge's folding in the formation of a tetrahedron}
While each hinge in the tetrahedron formation is a 3D structure of 4 layers hexagonal closest-packed lattice, here we analyze a 2D structure with only two rows of droplets as an approximation. Each row contains $N$ droplets. The initial radius of each droplet is $R_i$ and the final equilibrated radius of each droplet is $R_{1f}$ for the top row and $R_{2f}$ for the bottom row. The initial osmolarity is $C_{1i}$ for the top row, $C_{2i}$ for the bottom row, and the final equilibrated osmolarity is $C_f$ for both rows. The amount of solute in each droplet is conserved. This implies
\begin{equation}
\begin{aligned}
C_{1i}\frac{4}{3}\pi {R_i}^3 &= C_f \frac{4}{3}\pi {R_{1f}}^3, \\
C_{2i}\frac{4}{3}\pi {R_i}^3 &= C_f \frac{4}{3}\pi {R_{2f}}^3. \\
\end{aligned}
\end{equation}
The total amount of solution is also conserved:
\begin{equation}
2 \left( \frac{4}{3}\pi {R_i}^3 \right) = \frac{4}{3}\pi {R_{1f}}^3+\frac{4}{3}\pi {R_{2f}}^3.
\end{equation}
This yields  
\begin{equation}
\label{equation_R_C}
\begin{aligned}
R_{1f}&=\left ( \frac{2C_{1i}}{C_{1i}+C_{2i}} \right )^{\frac{1}{3}}R_i\\
R_{2f}&=\left ( \frac{2C_{2i}}{C_{1i}+C_{2i}} \right )^{\frac{1}{3}}R_i.\\
\end{aligned}
\end{equation}
\begin{figure}[h]
\begin{center}
\includegraphics[width=10.0cm]{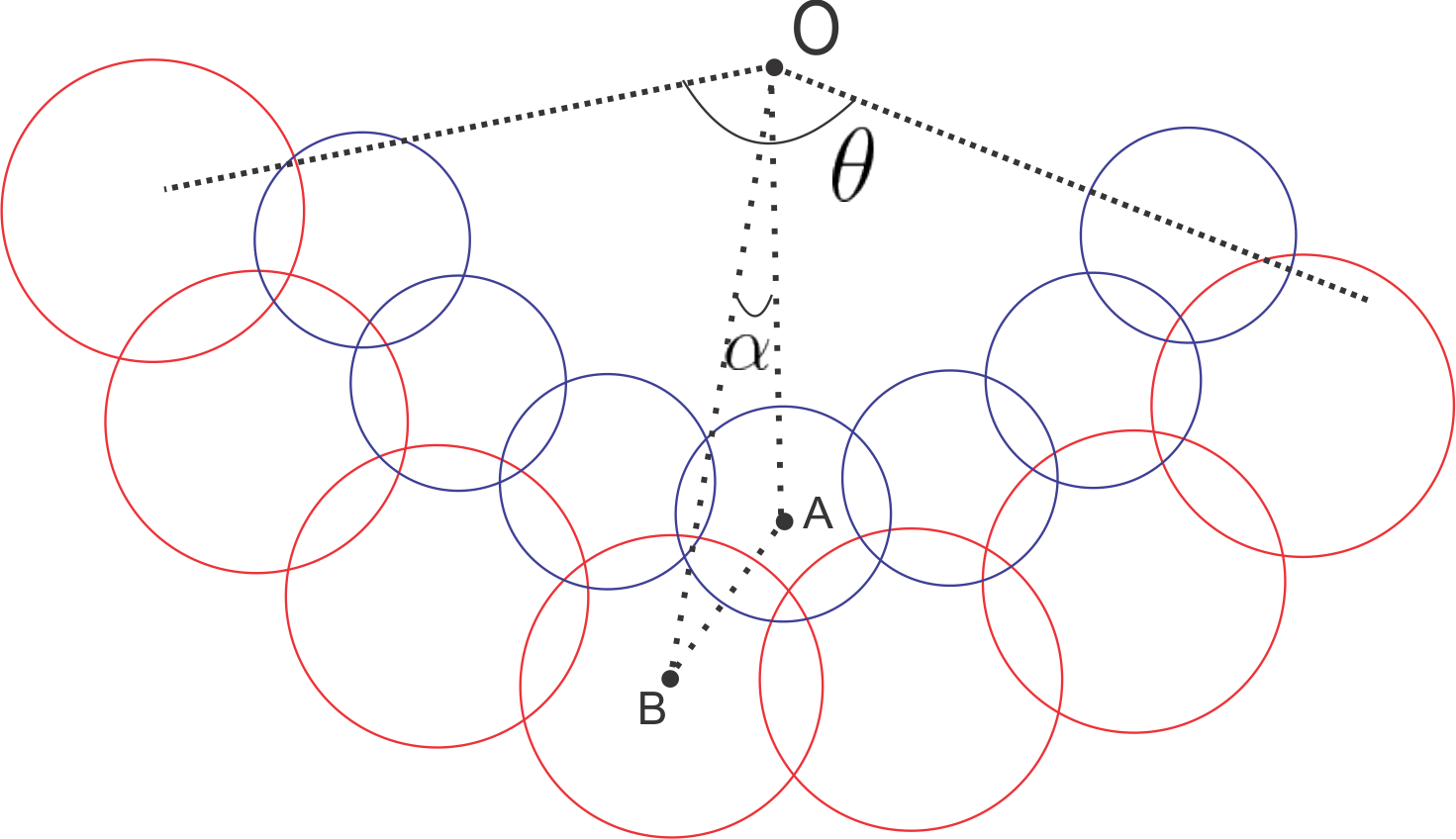}
\caption{ Schematic figure for the folding of a hinge.}
\label{tetrahedron_formation}
\end{center}
\end{figure}
To fold this hinge by an angle $\theta$, note that each row of droplets must have the same center of curvature, labelled as point $O$ in Fig.~\ref*{tetrahedron_formation}. The distance between two neighboring droplets in the top (inner) row is $2l R_{1f}$, the distance between two neighboring droplets, one from the top row and the other one from the bottom row, is $l \left ( R_{1f}+R_{2f}\right )$, and the distance between two neighboring droplets from the bottom row is $2 l R_{2f}$, where $l=0.8$ as shown in Fig.~\ref*{tetrahedron_formation}. In $\triangle OAB$, we have
\begin{equation}
\label{equation_triangle}
\begin{aligned}
|OA| &= \left(l R_{1f}\right)/\sin\alpha \\
|OB| &= \left(l R_{2f}\right)/\sin\alpha \\
|AB| &= l \left(R_{1f}+R_{2f}\right)\\
|AB|^2 &= |OA|^2+|OB|^2-2|OA||OB|\cos\alpha \\
\end{aligned}
\end{equation}
This yields for onset condition of folding this hinge by an angle $\theta = \left( 2N-1 \right)\alpha$
\begin{equation}
\label{equation_tetrahedron_formation}
\left(R_{1f}+R_{2f}\right)^2=\frac{R_{1f}^2+R_{2f}^2-2R_{1f}R_{2f}\cos\left(\frac{\theta}{2N-1}\right)}{\sin^2\left(\frac{\theta}{2N-1}\right)}
\end{equation}
By combining Eqs.~\ref{equation_R_C} and \ref{equation_tetrahedron_formation} we can numerically solve for $C_{2i}$ if given the value of $N$. For example, when $N=4$ and $C_{1i}=1.0$, we plot the onset value of $C_{2i}$ to fold the hinge by an angle $\theta$ as shown in Fig.~\ref*{tetrahedron_theta_C2i}.
\begin{figure}[h]
\begin{center}
\includegraphics[width=10.0cm]{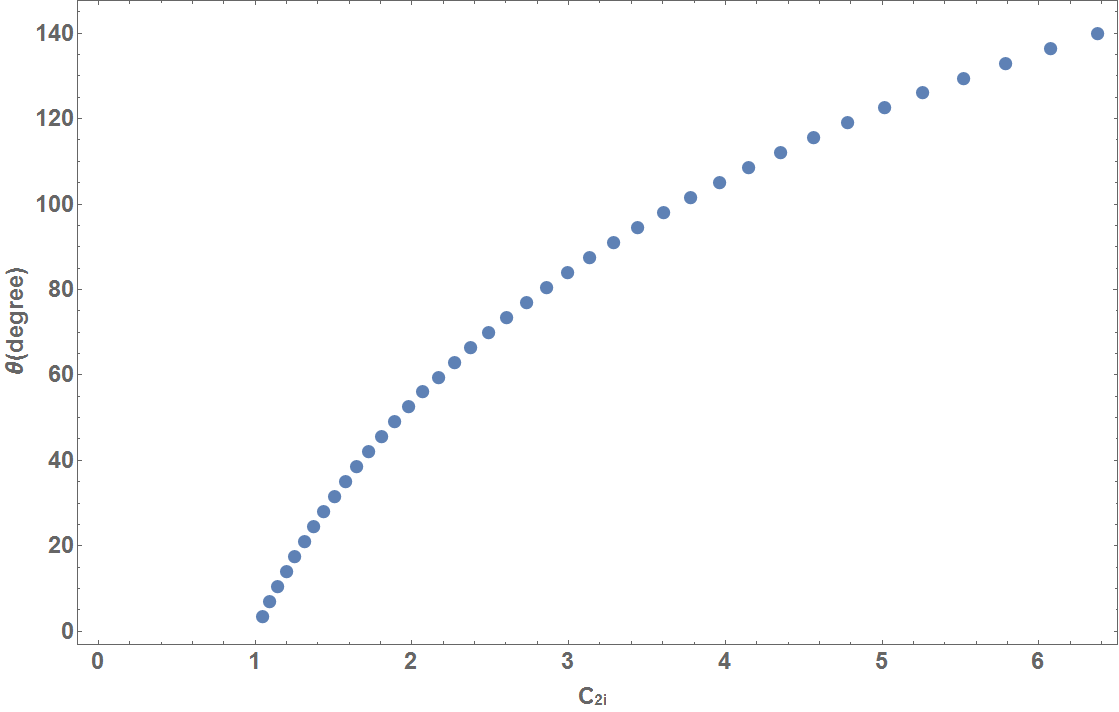}
\caption{The folding angle of a hinge $\theta$ as a function of the onset value of  $C_{2i}$, when $C_{1i}=1.0$ and $N=4$.}
\label{tetrahedron_theta_C2i}
\end{center}
\end{figure}
To form the tetrahedron, it requires $\theta=109.47^{\circ}$, then the onset value of $C_{2i}$ is $4.20$.}\\

\section{Details of analytical analysis of ring formation and subsequent buckling}
{We consider the ring formation of two rows with $N$ total droplets, as shown in Fig.~\ref*{S_ring_formation}. In this situation, similar to hinge folding, Eqs.~\ref{equation_R_C} and \ref{equation_triangle} are valid for the equilibrated configuration, with $\alpha=\frac{\pi}{N/2}$ in Eq.~\ref{equation_triangle}.
\begin{figure}[h]
\begin{center}
\includegraphics[width=10.0cm]{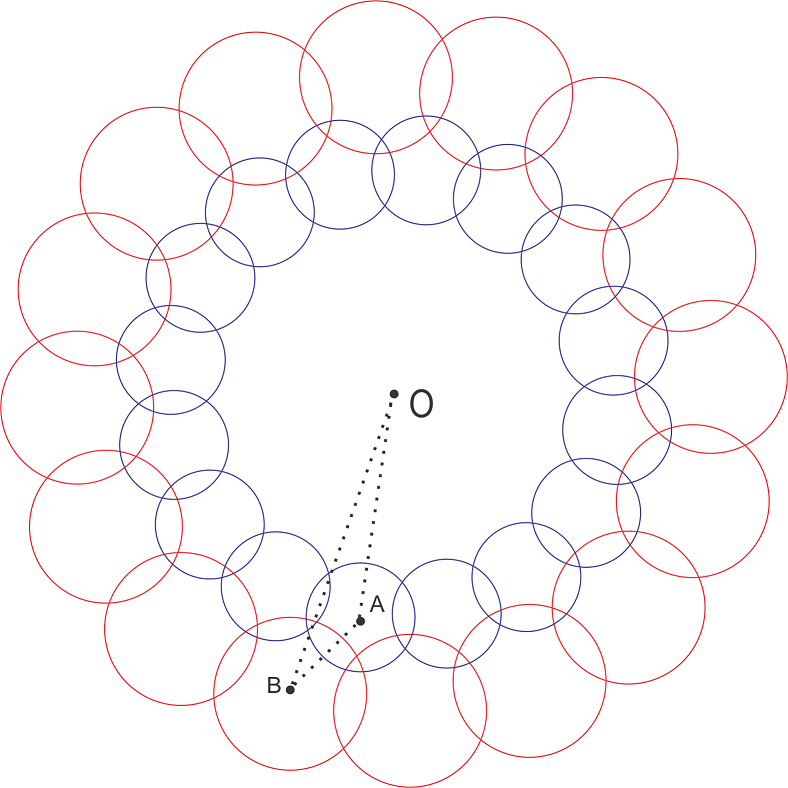}
\caption{Schematic figure for the onset of ring formation. }
\label{S_ring_formation}
\end{center}
\end{figure}
And the onset condition of ring formation is
\begin{equation}
\label{equation_ring_formation}
\left(R_{1f}+R_{2f}\right)^2=\frac{R_{1f}^2+R_{2f}^2-2R_{1f}R_{2f}\cos\left(\frac{\pi}{\left(N/2\right)}\right)}{\sin^2\left(\frac{\pi}{\left(N/2\right)}\right)}
\end{equation}
By combining Eqs.~\ref{equation_R_C} and \ref{equation_ring_formation}, if given the values of $C_{2i}$ and $N$, we can solve for $C_{1i}$ numerically.} 


\begin{figure}[h]
\begin{center}
\includegraphics[width=10.0cm]{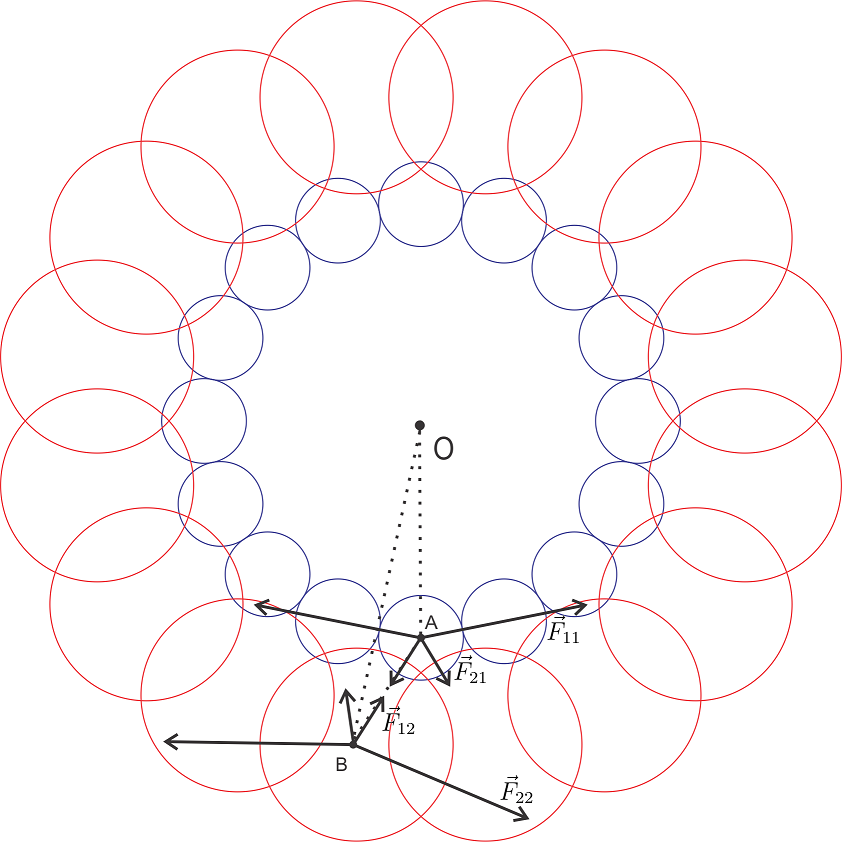}
\caption{Schematic figure for the onset of ring buckling. }
\label{S_ring_buckling}
\end{center}
\end{figure}

To determine the parameters for the onset of ring buckling where the inner droplets become separated, note that the distance between two neighboring droplets from the top(inner) row is $2 R_{1f}$, the distance between two neighboring droplets one from the top row and the other one from the bottom row is $l_{12}\left ( R_{1f}+R_{2f}\right )$, and the distance between two neighboring droplets from the bottom row is $ l_{22} \left(2 R_{2f}\right)$, as shown in Fig.~\ref*{S_ring_buckling}. For a droplet from the inner row, the force exerted by each of the two neighboring droplets from the same row is $| \vec{F}_{11}|=\left(1.0-l\right)\left(2R_{1f}\right)$, and the force exerted by each of the two neighboring droplets from the outer row is $| \vec{F}_{21}|=\left(l_{12}-l\right)\left(R_{1f}+R_{2f}\right)$, where $l=0.8$. Those four forces balance each other, yielding
\begin{equation}
2| \vec{F}_{11}|\sin\left(\frac{\pi}{\left(N/2\right)}\right)=2| \vec{F}_{21}|\frac{\sqrt{\left [ l_{12} \left ( R_{1f}+R_{2f} \right )\right ]^2-\left (l_{22} R_{2f}\right )^2}}{l_{12}\left ( R_{1f}+R_{2f}\right )}
\end{equation}
so that
\begin{equation}
\label{equation_l12_l22_1}
2 \left(1.0-l\right)\left(2R_{1f}\right) \sin\left(\frac{\pi}{\left(N/2\right)}\right)=2 \left(l_{12}-l\right)\left(R_{1f}+R_{2f}\right) \frac{\sqrt{\left [ l_{12} \left ( R_{1f}+R_{2f} \right )\right ]^2-\left (l_{22} R_{2f}\right )^2}}{l_{12}\left ( R_{1f}+R_{2f}\right )}
\end{equation}

For a droplet from the outer row, the force exerted by each of the two neighboring droplets from the same row is $| \vec{F}_{22}|=\left(l-l_{22}\right)\left(2R_{2f}\right)$ and the force exerted by each of the two neighboring droplets from the inner row is $| \vec{F}_{12}|=\left(l_{12}-l\right)\left(R_{1f}+R_{2f}\right)$, where $l=0.8$. Those four forces balance each other, yielding
\begin{equation}
2| \vec{F}_{22}|\sin\left(\frac{\pi}{\left(N/2\right)}\right)=2| \vec{F}_{12}|\frac{\sqrt{\left [ l_{12} \left ( R_{1f}+R_{2f} \right )\right ]^2-\left (R_{1f}\right )^2}}{l_{12}\left ( R_{1f}+R_{2f}\right )}
\end{equation}
so that
\begin{equation}
\label{equation_l12_l22_2}
2 \left(l-l_{22}\right)\left(2R_{2f}\right) \sin\left(\frac{\pi}{\left(N/2\right)}\right)=2 \left(l_{12}-l\right)\left(R_{1f}+R_{2f}\right) \frac{\sqrt{\left [ l_{12} \left ( R_{1f}+R_{2f} \right )\right ]^2-\left (R_{1f}\right )^2}}{l_{12}\left ( R_{1f}+R_{2f}\right )}
\end{equation}

We also have
\begin{equation}
\begin{aligned}
|OA| &= \left(R_{1f}\right)/\sin\left(\frac{\pi}{\left(N/2\right)}\right)\\
|OB| &= \left(l_{22} R_{2f}\right)/\sin\left(\frac{\pi}{\left(N/2\right)}\right)\\
|AB| &= l_{12} \left(R_{1f}+R_{2f}\right)\\
|AB|^2 &= |OA|^2+|OB|^2-2|OA||OB|\cos \left(\frac{\pi}{\left(N/2\right)}\right)\\
\end{aligned}
\end{equation}
The condition for the onset of ring buckling is thus
\begin{equation}
\label{equation_ring_buckling}
l_{12}^2\left(R_{1f}+R_{2f}\right)^2=\frac{R_{1f}^2+l_{22}^2R_{2f}^2-2l_{22}R_{1f}R_{2f}\cos\left(\frac{\pi}{\left(N/2\right)}\right)}{\sin^2\left(\frac{\pi}{\left(N/2\right)}\right)}.
\end{equation}
{By combining Eqs.~\ref{equation_R_C}, \ref{equation_l12_l22_1}, \ref{equation_l12_l22_2} and \ref{equation_ring_buckling}, we can numerically solve for $l_{12}$, $l_{22}$ and $C_{1i}$ for given values of $N$ and $C_{2i}$.}

\end{document}